\documentclass[aps,prl,twocolumn,preprintnumbers,
superscriptaddress,floatfix]{revtex4}

\usepackage{amssymb,multirow}
\usepackage{graphicx}

\begin{document}

\newcommand{\Tr}{\mbox{Tr\,}}
\newcommand{\beq}{\begin{equation}}
\newcommand{\eeq}{\end{equation}}
\newcommand{\bea}{\begin{eqnarray}}
\newcommand{\eea}{\end{eqnarray}}
\renewcommand{\Re}{\mbox{Re}\,}
\renewcommand{\Im}{\mbox{Im}\,}

\voffset 1cm

\newcommand\sect[1]{\emph{#1}---}

\preprint{
\begin{minipage}[t]{3in}
\begin{flushright} SHEP-08-17
\\[30pt]
\hphantom{.}
\end{flushright}
\end{minipage}
}

\title{The thermal phase transition in a QCD-like holographic model}

\author{Nick Evans and Ed Threlfall}

\affiliation{School of Physics and Astronomy, University of
Southampton,
Southampton, SO17 1BJ, UK \\
evans@phys.soton.ac.uk,ejt@phys.soton.ac.uk}

\begin{abstract}
\noindent We investigate the high temperature phase of a dilaton
flow deformation of the AdS/CFT Correspondence. We argue that
these geometries should be interpreted as the ${\cal N}=4$ gauge
theory perturbed by a SO(6) invariant scalar mass and that the
high-temperature phase is just the well-known AdS-Schwarzschild
solution. We compute, within supergravity, the resulting Hawking-Page phase transition
which in this model can be interpreted as a deconfining transition
in which the vev for the operator $Tr F^2$ dissolves. In the
presence of quarks the model also displays a simultaneous chiral
symmetry restoring transition with the Goldstone mode and other
quark bound states dissolving into the thermal bath.
\end{abstract}

\maketitle

\section{Introduction}

The simplest examples of non-supersymmetric deformations of the
AdS/CFT Correspondence \cite{Malda}-\cite{Gubser:1998bc} are those
in which the dilaton has some non-trivial profile in the radial
direction of the space
\cite{Kehagias:1999tr}-\cite{Constable:1999ch}. Since the dilaton
carries no R-charge the five-sphere is left intact. The presence
in the dilaton profile at large AdS radius, $r$, of a term of the
form $1/r^4$ indicates the presence of a vev for the operator $Tr
F^2$ in the ${\cal N}=4$ gauge theory. This operator is the F-term
of a chiral superfield so supersymmetry is manifestly broken.

Simple gravitational theories of this type have been shown to
generate confining behaviour in Wilson loops  in the dual gauge
theory and a discrete glueball spectrum
\cite{Gubser:1999pk,Ghoroku05}. Quarks have also been introduced
through D7 brane probes \cite{Karch}-\cite{Erdmenger:2007cm} and
chiral symmetry breaking in the pattern of QCD is induced
\cite{Babington,Ghoroku:2004sp}. These models are therefore a nice
toy model of a gauge theory that behaves in many respects like
QCD.

Given the successes of this model at zero temperature it is
interesting to investigate the finite temperature behaviour of the
solution. One expects a Hawking-Page type phase transition
\cite{Hawking:1982dh} where, when the temperature grows greater
than the perturbation, the vev for $Tr F^2$, a first order
transition would occur from the zero temperature solution with a
compact time dimension to a black hole geometry. The former has
free energy of order one whilst that of the latter is of order
$N^2$ -  we would be seeing deconfinement of the gauge degrees of
freedom \cite{Witten:1998qj,Witten:1998b}. It would be interesting
in addition to understand how chiral symmetry breaking behaves
through this transition.

Here we will seek dilaton flow black hole geometries that might
describe the high temperature phase of the gauge theory. In fact
though we will analytically show that there are no black hole
geometries with a non-trivial dilaton in five dimensional
supergravity.  The only candidate for the high temperature phase
of the dilaton flow geometry, at the supergravity level, is in
fact the normal AdS-Schwarzschild geometry. This is of course
identical to that describing the high T phase of the ${\cal N}=4$
gauge theory. We conclude that the vev for $Tr F^2$ is an induced
operator which dissolves at finite temperature. Taking this as our
assumption we compute the Hawking-Page type transition and show
that the behaviour matches the usual intuition discussed above, though we must caveat our analysis because we ignore potential string theory corrections to the highly-curved region of the low temperature phase. 

We learn that in the low temperature phase the vevs of the glue
and quark fields are temperature independent as are the glueball
and meson masses. This matches large N field theory arguments made
in \cite{Herzog0}. The finite temperature solution of the ${\cal
N}=4$ theory has also been studied vigorously \cite{SonStarinets}
including in the presence of quarks \cite{Erdmenger:2007cm}- all
of those results can now be seen to apply to the high temperature
phase of the dilaton flow geometry too. In particular the vev of
$Tr F^2$ and the chiral symmetry breaking quark condensate (at
zero quark mass) both switch off in the high T phase. The
Goldstone boson of the symmetry breaking becomes massive and
indeed melts into the thermal bath.

The above story though requires some additional explanation. In
the low T phase if the vev for $Tr F^2$ is an induced operator
what is the perturbation to the ${\cal N}=4$ gauge theory that is
breaking supersymmetry? We propose a story that might explain
this. Since supersymmetry is broken but SO(6)$_R$ preserved we
expect all other SO(6)$_R$ invariant operators to switch on. As
was argued in \cite{Gubser:1999pk} and we will discuss more below,
amongst these operators, at large $N$, is an SO(6) preserving
scalar mass term. This mass term is not described by a
supergravity field so is essentially invisible in the solutions -
there are examples of supersymmetric
\cite{Girardello:1999bd,Pilch:2000ue}and non-supersymmetric
\cite{Babington:2002qt} flows where this operator is present yet
invisible in terms of a supergravity field. In those geometries
the mass appears to be generated through the RG flow by a fermion
mass term that {\it is} described by a supergravity mode. Here a
simple D3 brane probe shows there is no explicit mass in the
dilaton flow geometry though and it is not in this class.

The SO(6)$_R$ invariant scalar mass is described by a string
rather than supergravity mode so one naively expects it to
describe a super-irrelevant source (if nothing else is there to
regenerate it). That is, if present, the source would be invisible
in the IR before growing sharply in the UV and dominating the
physics. Such a source term would show up as a sharp UV cut off,
so the only impact of that cut off in the low energy theory would
be the symmetries it imprinted. We propose that the dilaton flow
geometry describes the IR physics below such a UV cut off.

When the dual gauge theory is viewed as the ${\cal N}=4$ theory
with a scalar mass perturbation it seems a very natural theory to
study as a toy for QCD - it is a sensible non-supersymmetric,
strongly coupled gauge theory. Of course none of the gaugino
super-partners are decoupled from the strong dynamics so it is not
QCD.

Our results are also interesting as ten dimensional realizations
of the ``hard-wall" transitions explored in \cite{Herzog0} (the
introduction of a hard infra-red wall has long been used as a very
simplistic way of introducing a mass gap into the gauge theory
\cite{Polchinski:2001tt}). To make that connection stronger we
will begin by describing briefly the thermal transition in the
${\cal N}=4$ gauge theory with an SO(6) invariant scalar vev. This
should be the ultra-relevant operator that is described by the
string mode discussed above. The gravity dual of this theory is
precisely AdS$_5\times S^5$ with a hard IR cut off at some finite
radius - the cut off corresponds to the surface of a five-sphere
that the D3 branes have been spread evenly over. This case serves
as an example of how these scalar operators are invisible in the
supergravity solution.

Our main computation here is to show that there is no dilaton flow
black hole in five dimensional supergravity and to compute the
Hawking-Page transition to a pure AdS-like black hole in that
model. Finally we will briefly review the phase structure of the
non-supersymmetric gauge theory collating results from elsewhere
in the literature.

\section{A Hard wall - ${\cal N}=4$ SYM On Moduli Space}

We begin by creating a true AdS-dual of a hard wall model. The set
up is simple - one spreads the D3 branes at the origin ($r=0$) of
the usual AdS/CFT construction onto the surface of a five-sphere
centred at the origin. If a finite number of D3 branes are evenly
distributed then the configuration will preserve a discrete
sub-group of the SO(6) symmetry group. In the infinite N limit
where the distribution becomes smooth the SO(6) group is
preserved. In the gauge theory this configuration corresponds to
one where
\beq Tr \phi_i \phi_j \propto \delta_{ij} \eeq which is
an SO(6) singlet. The U(N) gauge theory is broken to a U(1)$^N$
theory that is non-interacting (all matter is in the adjoint so
uncharged) in the case of a finite number of D3 branes. As the
density increases the scalar vevs connecting adjacent sites become
small and one has a deconstructed model of a five dimensional U(1)
gauge theory living on the five sphere surface - again it is
non-interacting.

On the gravity side it is clear from a Gauss' law argument that
the geometry does not change (as in the case of a planet being
described by the Schwarzschild black hole metric). The dual is
just $AdS_5\times S^5$ with the usual four form (here $L^4 = 4 \pi
g_s N \alpha'^2$) \beq ds^2 = {u^2 \over L^2} dx_{4}^2 + {L^2
\over u^2} du^2 + L^2 d \Omega_5^2\eeq \beq C_{(4)} = {u^4 \over
L^4} dx^0\wedge..dx^3 \eeq except that the surface of the five
sphere of D3 branes acts as a cut off on the space (formally
within there is flat space since there are no sources).

The obvious candidate for the finite temperature version of the
theory is just AdS-Schwarzschild \cite{Witten:1998b}. Restricting
ourselves to a black hole background with temperature equal to
$\frac{u_h}{\pi L^2}$ and working with the usual Poincar\'e
coordinates
\begin{equation} \label{bh2}
ds^2 = {K(u) \over L^2} d \tau^2 + L^2 {du^2 \over K(u)} + {u^2
\over L^2} dx_{3}^2 + L^2 d \Omega_5^2
\end{equation} with
\begin{equation}
K(u) = u^2 - {u_h^4 \over u^2}
\end{equation}

We can seek a Hawking-Page transition by comparing the free energy
of AdS cut off at the radius of the D3 five sphere, $u_0$, and
with a compact time dimension versus that of the AdS Schwarzschild
solution cut off by either the radius of the five sphere or the
horizon whichever is largest.

This computation can be performed within the five dimensional
truncation of IIB supergravity on this space. The two five
dimensional metrics are then simply
\begin{equation} \label{ads}
ds^2 = {u^2 \over L^2} d \tau^2 + L^2 {du^2 \over u^2} + {u^2
\over L^2} dx_{3}^2
\end{equation}
\begin{equation} \label{bh}
ds^2 = {K(u) \over L^2} d \tau^2 + L^2 {du^2 \over K(u)} + {u^2
\over L^2} dx_{3}^2
\end{equation}
with no four forms. The comparison at this level is naively
precisely the computation of Herzog \cite{Herzog0} which we
briefly review. We will see shortly that in the full theory there
is an extra contribution to the computation.

The Euclidean action for either cut-off AdS or AdS-Schwarzschild is
\begin{equation}
S=-\frac{1}{4 \pi G_5} \int d^5 x \sqrt{g} \left ( \frac{1}{4} R+\frac{3}{L^2} \right )
\end{equation}

On-shell $R=-\frac{20}{L^2}$ for both backgrounds.

One must rescale the time coordinates so as to ensure that the
period of the time directions match at the cut off $\Lambda$ \cite{Witten:1998b}. One
then finds the action difference
\begin{equation}
S_{BH}-S_{AdS}=\frac{1}{2 G_5 u_h L^3} \left ( \int_{u_h}^{\Lambda} u^3 du -
\sqrt{K(\Lambda) \over \Lambda^2}\int_{u_0}^{\Lambda} u^3 dr \right
)
\end{equation}

Taking the limit $\Lambda \rightarrow \infty$ one obtains
$S_{BH}-S_{AdS}=\frac{1}{8 G_5 u_h L^3} \left ( u_0^4-\frac{1}{2} u_h^4 \right
)$.  This is Herzog's result.

This computation suggests that if $u_0>\sqrt[4]{\frac{1}{2}} u_h$
the black hole action is larger and a hard wall solution is
favoured - giving Herzog's transition temperature
$T_c=\frac{\sqrt[4]{2}u_0}{\pi L^2}$.

In fact the above computation can be interpreted as  a valid
result in the AdS-QCD approach, in which one constructs
phenomenological models without insisting that the holographic
physics is an exact realization of IIB SUGRA.  To study the actual
behaviour of the dual ${\cal N}=4$ gauge theory we must solve the
SUGRA problem in its entirety.  There is an additional term in the
action from the boundary between the flat spacetime within the
shell of D3 branes and the AdS geometry outside, which is the
difference between the Gibbons-Hawking term from each `side' of
the boundary (we thank Andreas Karch  and Steve Paik for pointing
this out to us). It is simplest to perform the calculation in the
full $10D$ geometry. The Gibbons-Hawking term is $\frac{1}{8 \pi
G_5} \int_{\Sigma} K d \Sigma$ where the integration is over the
boundary $\Sigma$ and $K$ is the trace of the second fundamental
form on the boundary. This is easily evaluated using the relation
\cite{Hawking:1977}

\begin{equation}
\int_{\Sigma} K d \Sigma= \frac{\partial}{\partial n} \int_{\Sigma} d \Sigma
\end{equation}

The normal derivative is evaluated by setting the  metric
coefficient of the radial holographic direction to unity (by means
of a coordinate transformation) and then differentiating with
respect to the radial direction.

The ten-dimensional flat space within the D3 brane shell has line
element  $ds^2= L^2 \left (d\tau^2 + dx_3^2 \right ) +
\frac{1}{L^2} \left (du^2 + u^2 d\Omega_5^2 \right )$ and thus
$\int_{\Sigma} d \Sigma=u^5$ up to constants, a multiple of the
five-sphere volume and four-space volume. Performing the normal
derivative one obtains a contribution of $5 u_0^4$.

The $AdS_5 \times S^5$ geometry outside the D3 brane  shell has
line element $ds^2= L^2 e^{2r} (d\tau^2 + dx_3^2)+\frac{1}{L^2}
\left ( dr^2+ d \Omega_5^2 \right )$ and thus $\int_{\Sigma} d
\Sigma=e^{4r}$ up to volume factors. Performing the normal
derivative and transforming back to the `$u$'-type coordinates one
finds a contribution of $4 u_0^4$ times the overall factors.

Including this in our computation we see that  it cancels out the
total action for the cut-off AdS spacetime entirely leaving

\begin{equation}
S_{BH}-S_{AdS}=\frac{1}{8 G_5 u_h L^3} \left ( -\frac{1}{2} u_h^4 \right )
\end{equation}

This implies that the Hawking-Page transition actually takes place
at any finite  temperature for the theory on its moduli space,
which is the same result as for the theory at the superconformal
point. Field theoretically this is because the temperature
generates a potential for the adjoint scalars of the gauge theory
which forces their vev to zero. The transition then naturally
occurs immediately above T=0.

\section{Dilaton Flow Geometries}

We now turn to constructing solutions of the supergravity
equations of motion with non-trivial dilaton flows.

\subsection{Five-dimensional action and equations of motion}

We will work in $\mathcal{N}=8$ SUGRA in five dimensions
\cite{Gunaydin:1984qu}-\cite{Gunaydin:1985cu} which is a
truncation of IIB string theory on $AdS_5 \times S^5$ and it is
known that any solution can be lifted to a complete ten
dimensional geometry. The 40 scalars which participate in the
superpotential will be set to zero (leaving a constant
superpotential which acts as a negative cosmological constant) and
we consider a solution with nontrivial dilaton and zero axion.

The effective five-dimensional action is (we use the normalization
conventions of \cite{Gubser:1999pk} and set $L \equiv 1$ for this section)
\begin{equation}
S=\frac{1}{4 \pi G_5} \int d^5 x \sqrt{-g} \left ( \frac{1}{4} R -
\frac{1}{8}g^{ab} \nabla_a \phi \nabla_b \phi +3 \right )
\end{equation}

The non-extremal background and the background with a nontrivial
scalar are both non-supersymmetric and we cannot apply the
technique of Killing spinor equations.  Instead we use symmetry to
constrain the form of the solutions.

We will make an ansatz, following the analysis of the ${\cal
N}=2^*$ gauge theory in \cite{Buchel}, of the form
\begin{equation}
ds_5^2= e^{2A} \left ( -e^{2B} dt^2 + dx_3^3 \right ) + dr^2
\end{equation}
The presence of $A$ allows the dilaton to have a non-trivial $r$
dependence and that of $B$ allows for non-zero temperature.

The field equations are
\begin{eqnarray}
\frac{1}{4} R_{ab} &=& \frac{1}{8} \partial_a \phi \partial_b \phi- g_{ab} \\
\nabla^2 \phi &=& 0
\end{eqnarray}

Using the linear combinations $\bar{A} \equiv A +\frac{1}{4}B$ and
$\bar{B}  \equiv \frac{\sqrt{3}}{4} B$ as in \cite{Buchel} these
equations take the form
\begin{eqnarray}
\phi''+ 4 \bar{A}' \phi' &=&0\\
\bar{B}''+4 \bar{A}' \bar{B}' &=&0\\
6 (\bar{A}')^2 &=& \frac{1}{4} (\phi')^2+ 2(\bar{B}')^2+6
\end{eqnarray}

The first two equations can be integrated to yield $\phi'= c_1
e^{-4 \bar{A}}$  and $\bar{B}'=c_2 e^{-4 \bar{A}}$.  We will see
later that the two solutions we concentrate on correspond to
setting either one or the other of these constants zero. Defining
$6 c_3^2 \equiv  ( \frac{1}{4} c_1^2+2c_2^2 )$ we obtain
\begin{equation}
(\bar{A}')^2=c_3^2 e^{-8 \bar{A}}+1
\end{equation}

These equations are analytically solvable with solutions
\begin{equation}
e^{4 \bar{A}} = { c_4^2 e^{8r} - c_3^2 \over 2 c_4 e^{4r} }
\end{equation}
\begin{equation} \label{Bbar}
\bar{B} = {c_2 \over 4 c_3} \ln \left( {c_4 e^{4r} - c_3 \over c_4
e^{4r} + c_3}\right) + B_0
\end{equation}
\begin{equation}
\phi = {c_1 \over 4 c_3} \ln \left( {c_4 e^{4r} - c_3 \over c_4
e^{4r} + c_3}\right) + \phi_0
\end{equation}
For any solution that returns to AdS asymptotically $B_0=0$ and
$\phi_0$ is the dilaton value in the AdS limit.

\subsection{Solution with no event horizon}

Let us first take the solution above and set $B \equiv 0$ to find
a zero temperature dilaton flow with manifest 4D Lorentz
invariance.   The solution can be recast (by setting $c_3 = c_4
\zeta$)  in the form
\begin{eqnarray}
e^{2A} &=&  \sqrt{c_4 \over 2} ~ \sqrt{e^{4r}- \zeta^2 e^{-4r}} \\
\phi&=&\sqrt{3 \over 2} \ln \left ( \frac{e^{4r}- \zeta}{e^{4r}+
\zeta} \right ) + \phi_0
\end{eqnarray}

To match to other results in the literature \cite{Ghoroku:2004sp}
one can rescale the $x_4$ coordinates to effectively set
$c_4=1/2$,
  set $2 u^2 = e^{2r}$ and $\zeta =- 4 u_0^4$.
Reinstating the five sphere and moving to string frame one arrives
at the 10D metric
\begin{equation}
ds^2 = e^{\phi/2} \left( {u^2 \over L^2} {\cal A}^2(u) dx_4^2 +
{L^2 \over u^2} du^2 + L^2 d \Omega_5^2 \right)
\end{equation} with
\begin{equation}
{\cal A}(u) = \left( 1 - \left ({u_0 \over u} \right )^8 \right)^{\frac{1}{4}},
~~~~~ e^\phi = \left( {(u/u_0)^4 + 1 \over (u/u_0)^4 - 1}
\right)^{\sqrt{3/2}}
\end{equation}

The four form is just that in pure AdS. This metric clearly
becomes $AdS_5 \times S^5$ at large $u$ and has a deformation
parameter $u_0^4$ which has dimension four and no R-charge - this
parameter is naturally identified with $Tr F^2$ in the gauge
theory. Since $Tr F^2$ is the F-term of a chiral superfield
supersymmetry is therefore broken in this gauge theory.

A crucial aspect of the geometry is that it is singular at $u=u_0$
with the dilaton blowing up. A singularity should be a source of
unease and we do not have a full explanation of it but we wish to
argue there are number of ideas that suggest such geometries are
worthy of study none the less. The presence of D3 branes in the
geometry do provide sources that, in some non-supersymmetric
configuration, might complete the geometry (compare to the hard
wall model where they account for the discontinuity between AdS
and flat space). The N=2$^*$ geometry \cite{Pilch:2000ue}is also
singular at a point where the effective gauge coupling diverges -
this geometry has been matched to the expected field theory
solution at a particular point on moduli space
\cite{Buchel:2000cn}. This model provides evidence that a
divergent gauge coupling can show up as a divergence in the
geometry.

Our real motivation for the continued study here though is the
phenomenological successes of the geometry. It has been shown to
be confining in \cite{Gubser:1999pk,Ghoroku05} and to break chiral
symmetries when quarks are introduced \cite{Ghoroku:2004sp} as we
will discuss below. In this sense we can think of it as a back
reacted hard wall with the correct properties to describe QCD-like
physics.

\subsection{Black hole geometry}

Let us now turn to finding the high temperature phase of the
dilaton flow geometry just explored.

To have a solution with a horizon we will choose constants such
that the function $B$ goes as a constant plus $\ln r$ near $r=0$
($\bar{B} \sim \sqrt{3}/4 \ln r$). From (\ref{Bbar}) this gives
the constraints
\begin{equation}
c_2=\sqrt{3} c_3, \hspace{1cm} c_4=c_3
\end{equation}

At this point we note that, with the definition of $c_3$ ($6 c_3^2
\equiv  ( \frac{1}{4} c_1^2+2c_2^2 )$),  $c_1$ vanishes so the
dilaton profile in the non-extremal solution is just a constant.
This result is a simple example of a `no hair' theorem.

We have learnt that there is no black hole solution with a
radially dependent dilaton. This means there is not a gravity dual
of a high temperature theory with $Tr F^2$ switched on. One's
immediate response is to become worried that if the ${\cal N}=4$
gauge theory cannot be perturbed by a vev for $Tr F^2$ at finite
temperature then the zero temperature model is suspect. In fact
though we believe it is telling us that the vev for $Tr F^2$ is
not the fundamental perturbation but an operator induced by the
dynamics. We will discuss what the true perturbation might be in
the next section.

We are left with a unique black hole solution which in the
unbarred quantities is thus
\begin{eqnarray}
e^{2A} &=& \frac{c_4}{2}  \left ( e^{2r}+e^{-2r} \right ) \\
B &=& \ln \left ( \frac{e^{4r}-1}{e^{4r}+1} \right )\\
\phi &=&  \phi_0
\end{eqnarray}
This geometry should describe the high temperature phase of the
dilaton flow theory. In fact by rescaling $x_4$ to set $c_4=2 u_h^2$
and defining $e^{2r}=\frac{u^2+\sqrt{u^4-u_h^4}}{u_h^2}$ (which are
asymptotically the same choices as for the dilaton flow geometry)
this solution can be reduced to the usual Poincar\'e coordinate form
of five-dimensional AdS-Schwarzschild (\ref{bh2}) with Hawking
temperature $T_H=\frac{u_h}{\pi}$. We conclude that the scalar
mass deformed gauge theory shares the same supergravity
description at finite temperature as the unperturbed ${\cal N}=4$
gauge theory!  A similar conclusion was reached in \cite{SangjinSing} in the context of adding a dilaton to the hard wall model.

\section{The Origin Of Supersymmetry Breaking}

We now turn to the question of the origin of supersymmetry
breaking in the dilaton flow geometry if $Tr F^2$ is an induced
operator as it appears it must be from the above analysis.

Since supersymmetry is broken, yet SO(6)$_R$ preserved, in the T=0
geometry, we expect all SO(6) invariant operators to be present.
Amongst these SO(6) invariant operators is an equal mass for each
of the six scalar fields - one would expect the scalar masses to
rise to the scale of the supersymmetry breaking scale. Such an
SO(6) invariant mass is invisible in the supergravity solution for
the same reasons as the SO(6) scalar vev operator discussed above.
A frequently argued interpretation of the fact that this source is
not described by a supergravity mode is that it is a
super-irrelevant operator. The vev for the scalar operator
discussed in the hard wall model above would then be
super-relevant; that is it would have no impact on the UV of the
theory until suddenly at some point in the IR it would dominate it
- this can certainly be matched to it's appearance as a sharp IR
cut off on the geometry. In this language one would expect the
mass term to show up as a sharp cut off on the space at some large
radius or UV scale. Below that scale it would naively have no
impact on the dynamics. This is not quite true though because it
would define the symmetries of the theory below that UV cut off
and in particular leave a non-supersymmetric but SO(6) invariant
flow at lower energies. The dilaton flow geometry is the natural
candidate for this flow. In this interpretation one should cut off
the dilaton flow at some point in the UV, although this point
could presumably be set at an arbitrarily high scale. In analogy
with the hard wall model the point where the cut off appears would
be undetectable in the low energy flow.

The above seems a consistent interpretation but the supersymmetric
N=1$^*$ \cite{Girardello:1999bd} and N=$2^*$ \cite{Pilch:2000ue}
theories suggest a more complicated story is also possible. In
those theories precisely the scalar mass discussed here is present
in the Lagrangian of the gauge theory yet no supergravity mode
directly represents it in the supergravity duals. The scalar mass
must, by supersymmetry, be present and tied to the fermion masses
(in superspace there is only the one mass parameter) which are
described by supergravity fields. The flows for these fields show
the mass term to be relevant. These are therefore examples of
theories with a relevant scalar mass present but no explicit dual
operator to indicate it. The Yang Mills$^*$ geometry
\cite{Babington:2002qt} is a non-supersymmetric example - a
fermion mass term is introduced there yet the potential for a D3
probe shows there to be a scalar mass present too.  This is the
simple way for us to test whether there is a scalar mass present
here - we look at the potential for moving a D3 probe in the
transverse directions of the dilaton flow geometry. The result is
clear on simple dimensional grounds - the asymptotic potential for
the D3 motion must go like $u_0^4$ (note the fourth power is the
lowest to occur in the metric) to some positive power so that it
vanishes when $u_0$ does. We can only have a $u$ dependent term of
the form $V \sim u_0^8/u^4$ - there is no $m^2 u^2$ term and so no
explicit mass. The full D3 potential is given by \beq V ~\sim~ u^4
{\cal A}^4 - u^4 ~\sim~ - {u_0^8 \over u^4} \eeq which is stable.

Presumably in the theories with fermion masses the scalar mass is
continually regenerated through the RG flow whilst if only the
scalar mass is present it flows to zero in the IR. We conclude
that in the dilaton flow geometry the scalar mass would indeed
only be visible through a sharp UV cut off as discussed above.
This seems a consistent interpretation to us and with this in mind
we will go on to analyze the high temperature transition in the
dilaton flow geometry.

Note this geometry is also closely related to those of Constable
and Myers \cite{Constable:1999ch} which have in addition
non-trivial $u$ dependence in the four form - the existence of
this larger class of geometries suggest that there are multiple
SO(6) invariant string modes that are invisible in the
supergravity and that determine the dynamics. In the field theory
one can imagine higher dimension operators and so forth that could
play a role. These geometries typically also show confinement and
chiral symmetry breaking though \cite{Babington}.

\section{Thermodynamic computation}

One way to test our assertion that AdS-Schwarzschild  is the high
temperature phase of the non-supersymmetric deformation of the
${\cal N}=4$ gauge theory is to check the Hawking-Page phase
transition makes sense.  We will compute the Euclidean action for
both solutions,  specifying  a black hole horizon at $u=u_h$ and a
dilaton flow singularity at $u=u_0$.

To make the comparison fair we must set the parameter $c_4$ equal
in the two geometries so they have the same large-$r$ AdS limit.
We will perform the calculation in the Schwarzschild-type
coordinates,  rescaling the Euclidean time coordinate for the
dilaton flow geometry as for our hard-wall calculation.  Both
geometries asymptote to $AdS_5$ so we can set the same UV cut-off
$\Lambda$ in both cases, before taking the limit $\Lambda
\rightarrow \infty$.

Our interpretation above that the dilaton flow geometry is the IR
theory below some UV cut off associated with the presence of a
scalar mass means that formally we should keep the UV cut off
fixed. We can though imagine that that scale is arbitrarily high.
In any case we will give the result for arbitrary $\Lambda$ below.

The Euclidean action density per unit spatial volume for the black hole solution is
\begin{equation}
S_{BH}=- \frac{1}{4 \pi G_5} \int_0^{\frac{\pi L^2}{u_h}} d \tau \int_{u_h}^{\Lambda} \sqrt{-g}  \left (
\frac{1}{4} R+\frac{3}{L^2} \right ) dr
\end{equation}

The trace of the Einstein equation gives $R=-\frac{20}{L^2}$ so
\begin{equation}
S_{BH}=\frac{1}{2 G_5 u_h L^3} \int_{u_h}^{\Lambda} u^3 du \; = \; \frac{1}{8 G_5 u_h L^3} \left ( \Lambda^4-u_h^4 \right )
\end{equation}

The Euclidean action density per unit spatial volume for the
dilaton flow solution is, having used the trace of the equation of
motion to remove the scalar gradient term and allowing for the rescaling of Euclidean time, simply
\begin{equation}
S_{DF}=\frac{1}{2 G_5 L^2} \int_0^{\frac{\pi L^2}{u_h} \sqrt{1-\frac{u_h^4}{\Lambda^4}}} d \tau
\int_{u_0}^{\Lambda} \sqrt{-g} \; dr
\end{equation}

This is
\begin{eqnarray}
S_{DF}&=&\frac{1}{2 G_5 u_h L^3} \sqrt{1-\frac{u_h^4}{\Lambda^4}} \int_{u_0}^{\Lambda}
 \left ( u^3 - \frac{u_0^8}{u^5} \right ) dr \nonumber \\& = &
\frac{1}{8 G_5 u_h L^3} \sqrt{1-\frac{u_h^4}{\Lambda^4}} \left ( \Lambda^4+ \frac{u_0^8}{\Lambda^4}-2 u_0^4
\right )
\end{eqnarray}

Hence, in the $\Lambda \rightarrow \infty$ limit,  the difference
in the actions is simply
\begin{equation}
S_{BH}-S_{DF}= \frac{1}{16 G_5 u_h L^3} \left ( 4 u_0^4-u_h^4 \right )
\end{equation}

For a deformation scale $u_0 > \frac{u_h}{\sqrt{2}}$ the dilaton
flow solution is thermodynamically  favoured whereas for a
deformation scale $u_0<\frac{u_h}{\sqrt{2}}$ the AdS-Schwarzschild
solution is favoured.   The transition temperature is $T_c=\frac{\sqrt{2} u_0}{\pi L^2}$.

The phase transition appears to make complete sense with our interpretation
of the high and low temperature phases. For temperatures below the
value of the supersymmetry breaking scale $u_0$ (up to a factor of
order unity) the non-supersymmetric SO(6) invariant scalar mass
deformed ${\cal N}=4$ gauge theory is described by the dilaton
flow geometry with an induced vev for $Tr F^2$. As the temperature
passes through the supersymmetry breaking scale there is a
transition to the deconfined plasma described by AdS-Schwarzschild
- here the vev of $Tr F^2$ is zero.

It is important to stress though that the computation above may not be complete. We saw
above that when the hard wall model was converted into a full string geometry a Gibbons-Hawking term appeared at the IR singularity that played a crucial role in the thermodynamics
of the ${\cal N}=4$ theory on moduli space. One must worry that a similar surface term might
appear if the dilaton flow geometry were completed to a fuller string theoretic understanding.
In addition the negative term in our action computation is dominated near the singularity and
might also change were the singularity resolved in someway. Within supergravity, our only available
tool, it is hard to see how to address these complications - it is encouraging though
that the calculation as is does agree with expected field theory intuition and the naive hard wall
geometry as applied to QCD. It is interesting to compare this case to the AdS-QCD models of \cite{Gursoy}
in which the action computation is dominated away from the IR singularity - those theories may have
better control.
 
Another interesting point is that the supergravity computation naively suggests the dilaton flow
geometry's action is lower than that of AdS! Our discussion of the origin of supersymmetry
breaking though suggests this is not a correct comparison - we have argued that the geometry
only applies below some UV cut off corresponding to the scale where a super-relevant scalar mass
becomes important. It is an artefact of supergravity that this cut off is invisible in the IR.
Above that cut off the presence of a non-normalizable mode would make the dilaton
flow geometry's action much greater than the case of AdS extended to infinity. We do not expect the
${\cal N}=4$ theory to spontaneously break supersymmetry. This does not affect
our computation since both the AdS Schwarzschild black hole and the dilaton flow geometry share the
same UV (at least if the cut off is far enough into the UV) and hence would see the same UV cut
off physics. 

\section{Aspects of the Phase Transition}

The identification of AdS-Schwarzschild as the high T phase of a
non-supersymmetric theory and the dilaton flow as the low
temperature phase of that same theory is quite remarkable.   Our
findings go some way towards explaining why the AdS-Schwarzschild
gravity dual is a reasonable toy model of high-temperature QCD
whereas the zero-temperature supersymmetric D3-D7 model is quite
unlike low-temperature QCD (it is conformal, for example, when
the quarks are massless). Both of the geometries have been studied
in detail already in the literature, including in the presence of
quarks, and we can look at a number of properties of the
transition.

Below the critical temperature, the core of the dilaton flow
geometry is repulsive to strings. The result is that a Wilson line
calculation shows there to be a linear quark-antiquark potential
as the string falls towards $r=0$ before settling a little away
from the singularity at $u_0$ \cite{Gubser:1999pk,Ghoroku:2004sp}.
In keeping with the implied confinement there is a discrete
spectrum of glueballs which are the eigenmodes of the Klein-Gordon
equation on the geometry for the usual plane-wave ($\propto e^{i k
\cdot x_4}$) ans\"atze. Table I shows the masses of the lowest
five scalar glueball states, in units of $u_0/L^2$ (we can
reproduce the field equations for these fluctuations in
\cite{Gubser:1999pk} but disagree on the numerical values for the
masses) \footnote{The dilaton equation of motion is given by $\partial_u ( u^5 {\cal A}^4 \partial_u \phi) + u {\cal A}^2 M^2 \phi = 0$ (here we write $u$ in units of $u_0$ and rescale $x_4$ so that factors of $L,u_0$ are common
to the metric). The UV solutions take the form $\phi \sim c_1 + c_2/u^4$ with the latter
being required for a glueball fluctuation. In the IR the equation can be recast in
Schr\"odinger form - we write $u = 1 + z$, then change coordinates to $y$ such that
$dy/dz = 1/(8 z)^{1/4}$, and finally write $\phi = u v$ with ${1 \over v}
{d v \over dy} = -3/(8z)^{3/4}$. The potential is then of the form $V = -{1 \over 4 y^2}$.
Such a potential is of the limiting form that possesses a discrete spectrum bounded from below
(see \cite{Petrini} or chapter 5 of \cite{Landau}). The IR solutions, written in the original $u$
coordinates are of the form $\phi \sim c_3 + c_4 \ln{(u-1)}$ - the former are the physical
solutions, the latter blow up and are therefore inconsistent with linearization.
All of this is in complete agreement with the analytic discussion in \cite{Gubser:1999pk} and
we can numerically shoot from both the IR and UV solutions to find the values of $M$ for
which solutions match to the required UV and IR boundary conditions. We disagree with \cite{Gubser:1999pk}
on the numerical values of these solutions though.}

\begin{centering}
\begin{tabular}[t]{|c|ccccc|}
\hline $n$ & 1 & 2 & 3 & 4 & 5 \\ \hline
$M_n$ & 4.1 & 7.2 & 10.2 & 13.2 & 16.2 \\
\hline
\end{tabular}

\noindent Table I: Lowest five glueball masses in the zero
temperature dilaton flow geometry in units of the deformation
scale $u_0/L^2$.
\end{centering}

\begin{centering}
\begin{tabular}[t]{|c|c|}
\hline
$n$&$\omega_n$\\
\hline
1& $\pm$ 3.119452 - 2.746676 i \\
2& $\pm$ 5.169521 - 4.763570 i \\
3& $\pm$ 7.187931 - 6.769565 i \\
4& $\pm$ 9.197199 - 8.772481 i \\
5& $\pm$ 11.202676 - 10.774162 i \\
\hline
\end{tabular}

\noindent Table II: Lowest five glueball quasi-normal modes in the
AdS-Schwarzschild geometry in units of $\frac{u_h}{L^2}$.
\end{centering}

Above the critical temperature, the theory is in a deconfined
phase.   There is no longer a spectrum of glueball normal modes,
rather the gravity dual admits a spectrum of unstable quasinormal
modes which was calculated in \cite{Starinets}.  The field theory
interpretation of the quasinormal spectrum is to give the mass and
decay width for a glueball excitation embedded in a thermal bath
of SYM plasma.  The finite decay timescale can be viewed as the
timescale for the `melting' of the glueball state.  The breaking
of Lorentz symmetry means there is a nontrivial dispersion
relation $\omega(k)$ for a scalar glueball excitation
\cite{Starinets}.  Table II shows the lowest five quasinormal
frequencies, which are measured in units of $\frac{u_h}{L^2}$
which is $\propto T$ - the natural scale of the lowest quasinormal
frequency is the temperature.

Flavour degrees of freedom can be included by embedding D7 branes
in each of the geometries discussed
\cite{Polchinski,Bertolini:2001qa,Erdmenger:2007cm} - the results
to date in these geometries use the probe or quenched limit
\cite{Karch}. The asymptotic profile of the D7 encodes the
relationship between the hard quark mass $m_q$ and the expectation
value of the quark condensate $\langle \bar{q}q \rangle$
\cite{Babington}.  Mesonic modes are dual to fluctuations derived
from the DBI action of the D7 \cite{Mateos}.

Below the critical temperature we find the probe D7 always wants
to lie outside the deformation scale of the dilaton flow geometry
for the embeddings of physical interest (usefully avoiding the
singular region of the geometry).  One finds there is a nonzero
value of $\langle \bar{q}q \rangle = 1.51 u_0^3$ for zero $m_q$,
indicating spontaneous breaking of a U(1)$_R$ symmetry of the
model (this symmetry is analogous to the axial U(1)$_A$ of QCD) -
this is a nice model of QCD-like behaviour since the dynamics of
the quark condensate generation is included even if the full
non-abelian chiral symmetry breaking is not present. There are
discrete spectra corresponding to pion-like and sigma-like scalar
excitations \cite{Ghoroku05}. The lowest pion-like state is
massless for $m_q=0$ and its mass grows in accordance with the
Gell-Mann-Oakes-Renner relation for small $m_q$.  In addition
there is a tower of massive vector meson excitations dual to the
Maxwell field on the D7 worldvolume.  The numerical values for all
these meson masses tend to the no-deformation result (equation
(3.19) in \cite{Mateos}) for large $m_q$, that is $M \sim 2
\sqrt{2} m_q/L^2$ - we list them in Table III and display them in
Figure I in units of $u_0$.

\begin{centering}
\begin{tabular}[t]{|c|c|c|c|}
\hline
$m_q$&$M_{\pi} L^2$&$M_{\sigma}L^2$&$M_{vector}L^2$\\
\hline
0.10 & 0.7 & 3.1 & 2.9\\
0.50 & 1.9 & 3.5 & 3.3\\
1.00 & 3.1 & 4.1 & 3.9\\
2.00 & 5.7 & 6.0 & 6.0\\
3.00 & 8.5 & 8.6 & 8.6\\
4.00 & 11.3 & 11.4 & 11.4\\
\hline
\end{tabular}

\noindent Table III: the mass of the pion, sigma and rho meson
modes as a function of the quark mass in the low temperature
dilaton flow geometry - all in units of $u_0$.
\end{centering}

\begin{centering}
\begin{figure}[h]
\begin{centering}
\includegraphics[width=80mm]{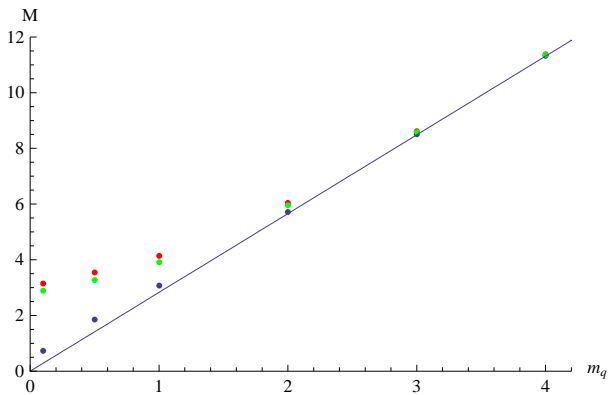}
\end{centering}
\caption{\small{Pion (blue), sigma (red) and vector (green) masses
as a function of quark mass - all in units of $u_0$.  The line
shows the large-$m_q$ limit.}}
\end{figure}
\end{centering}

Above the critical temperature the physically relevant D7
embeddings in the black hole geometry
\cite{Babington,Kirsch:2004km,Apreda,Mateos1,Mateos2,Albash} give
$\langle \bar{q}q \rangle=0$ for $m_q=0$ - there is no chiral
symmetry breaking and hence no pion-like meson. The D7 can either
end on the black hole horizon (small $m_q$) or for large enough
$m_q$ it has sufficient tension to support itself away from the
black hole - there is a first-order phase transition in the
behaviour of (quenched) quark matter as one raises the quark mass
in the plasma background. In the former case there are quasinormal
spectra representing the melting of scalar and vector mesonic
excitations in the hot background \cite{Zamaklar,Hoyos,Myers}. In
the latter case there are discrete spectra of scalar and vector
meson masses with scale set by $m_q$ which tend to the values in
\cite{Mateos} as $m_q \gg T$. Our transition behaves in the same
way as long as the quarks are sufficiently light ($m_q < 0.92
\frac{\sqrt{\lambda} T}{2}$) \cite{Hoyos}.  In the undeformed
theory this bound implies that the mesons melt once the
temperature of the background becomes of order the meson mass
since the meson masses are $\sim \frac{m_q}{\sqrt{\lambda}}$.  In
our case the pion-like meson is an exception to this - one can
have a massless pion that does not `melt' until the background
reaches some finite temperature.

\begin{centering}
\begin{tabular}[t]{|c|c|}
\hline
$n$&$\omega_n$\\
\hline
1& $\pm$ 2.1988 - 1.7595 i \\
2& $\pm$ 4.2119 - 3.7749 i \\
3& $\pm$ 6.2155 - 5.7773 i \\
4& $\pm$ 8.2172 - 7.7781 i \\
5& $\pm$ 10.2181- 9.7785 i \\
\hline
\end{tabular}

\noindent Table IV: the scalar mesonic quasinormal frequencies in
the high T phase ($m_q=0$) - in units of $\frac{u_h}{L^2}$.
\end{centering}

There are recent results concerning a lower bound for the ratio of
viscosity to entropy density of a strongly-coupled field theory
\cite{Policastro:2001yc}, $\frac{\eta}{s} \geq \frac{1}{4 \pi}$,
where the equality is for the deconfined phase of $\mathcal{N}=4$
SYM theory (for a review see \cite{SonStarinets}).  Our findings
show that strict equality also applies to certain
non-supersymmetric theories in their deconfined phase, physically
due to the universality of this phase in the large-$N$ limit.

We can perform an estimate of the deconfinement temperature in our
model.   The mass of the lowest-lying vector state for zero quark
mass can be compared to the mass of the $\rho$ meson,
experimentally 776MeV.  The vector mass is $2.80 u_0/L^2$ and the
deconfinement temperature is $T_c=\frac{\sqrt{2} u_0}{\pi L^2}$.
This gives an estimate for the deconfinement temperature of $T_c
\sim 124$ MeV.  This is very similar to the estimate produced by
the `hard-wall' model \cite{Herzog0} and is somewhat low compared
to real QCD.

\section{Conclusion}

We have found the finite-temperature version of a chiral symmetry
breaking dilaton flow.  Performing a calculation at the level of classical supergravity we found that for a temperature
sufficiently high that the black hole radius is greater than the
deformation scale, the geometry undergoes a first-order phase
transition to the AdS-Schwarzschild geometry.  From the
gauge-theory perspective, we have argued that the dilaton flow
solution results from turning on a $SO(6)$ invariant scalar mass
deformation.  This deformation is not described by a SUGRA mode
and is probably present only as a UV cut off that determines the
symmetries of the IR theory, but one does observe that the
non-trivial dilaton profile describes a running coupling and the
operator $Tr F^2$ is `induced'. We have found that at high
temperature this does not happen and $Tr F^2$ remains zero.
Incorporating other results already in the literature one can see
that the transition corresponds not just to a deconfinement
transition but also a simultaneous chiral symmetry restoration
transition if (quenched) quarks are introduced into the theory. We
believe this is the simplest, four dimensional, AdS/CFT derived
caricature of a QCD-like theory.

\vspace{1cm}

\noindent {\bf Acknowledgements:} ET would like to thank STFC for
his studentship funding. We are grateful to Johanna Erdmenger,
Ingo Kirsch and Kazuo Ghoroku for comments on the manuscript.

\end{document}